\documentclass[fleqn,twoside]{article}
\usepackage{espcrc2}

\usepackage{graphicx}
\usepackage[figuresright]{rotating}

\newcommand{\AmS}{{\protect\the\textfont2
  A\kern-.1667em\lower.5ex\hbox{M}\kern-.125emS}}
\def\ltsim{\hbox{\raise 2pt \hbox {$<$} \kern-1.1em \lower 4pt \hbox {$\sim$}}}
\def\ltapprox{\hbox{\raise 2pt \hbox {$<$} \kern-1.1em \lower 5pt \hbox 
{$\approx$}}}
\def\gtsim{\hbox{\raise 2pt \hbox {$>$} \kern-1.1em \lower 4pt \hbox {$\sim$}}}
\def\gtapprox{\hbox{\raise 2pt \hbox {$>$} \kern-1.1em \lower 5pt \hbox 
{$\approx$}}}
\def\arcsec{$^{\prime\prime}$}
\def\arcmin{$^{\prime}$}

\title{Diffuse radio emission from the Intracluster medium}

\author{L. Feretti\address{Istituto di Radioastronomia CNR/INAF\\
     Via P. Gobetti 101, 40129 Bologna, Italy, {\it lferetti@ira.cnr.it}},
        C. Burigana\address{Istituto di Astrofisica Spaziale CNR/INAF\\
     Via P. Gobetti 101, 40129 Bologna, Italy, {\it burigana@bo.iasf.cnr.it}},
        T.A. En{\ss}lin\address{Max-Planck Institut f\"ur Astrophysik\\
     Karl-Schwarzschild-Str. 1, Postfach 1317, 85741 Garching, Germany, {\it ensslin@mpa-garching.mpg.de}}
}       
\begin{document}

\begin{abstract}
An important aspect of the radio emission from galaxy
clusters is represented by the diffuse radio sources associated
with the intracluster medium: radio halos, relics and mini-halos.  
The radio halos and relics are indicators of cluster mergers, 
whereas mini-halos are detected at the center of cooling core clusters.
SKA will dramatically improve the knowledge of these sources,
thanks to the detection of new objects, and to detailed studies
of their spectra and polarized emission. SKA will also provide the
opportunity to investigate the presence of halos produced by 
radiation scattered by a powerful radio galaxy at the cluster centers.
\vspace{1pc}
\end{abstract}

\maketitle

\section{INTRODUCTION}

The main component of the intracluster medium (ICM) in  clusters of
galaxies is represented by the X-ray emitting thermal plasma, which
amounts to about 15-18\% of the total gravitating mass of a cluster
and consists of particles with energies of several keV.  In addition,
a fraction of galaxy clusters exhibit large-scale radio sources,
which have no optical counterpart and no obvious connection to the
cluster galaxies, and are therefore associated with the ICM.  Their
emission, of synchrotron origin,  demonstrates the existence of
relativistic electrons with energies of $\sim$ GeV in $\sim$ $\mu$G
magnetic fields.

\begin{figure}
\includegraphics[height=14pc]{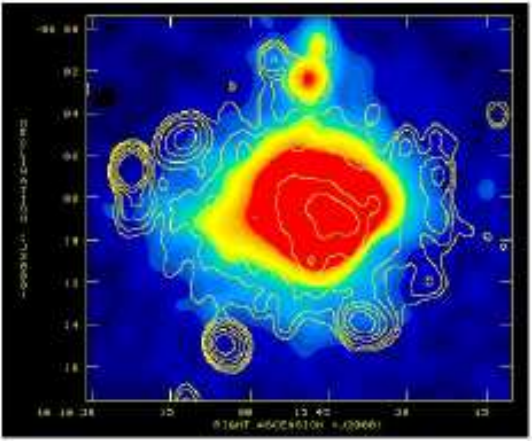}
\caption
{The cluster A2163 in radio and X-ray. The contours represent the
radio emission in A2163 at 20 cm, showing an extended radio halo (from
\cite{Fer01}).  The color scale represents the ROSAT X-ray
emission. The extended irregular X-ray structure indicates the
presence of a recent cluster merger.
\label{a2163rx}
}
\end{figure}

\begin{figure}
\includegraphics[scale=0.3]{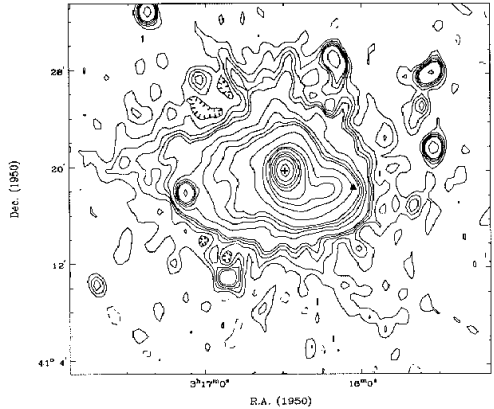}
\caption
{Radio contour map of the mini halo in the Perseus 
cluster, obtained at 92 cm with the Westerbork Synthesis Radio Telescope
at a resolution of
51\arcsec $\times$ 77\arcsec (RA$\times$DEC). 
The cross indicates the position of NGC1275, 
the triangle marks the position of NGC1272. The mini--halo
size in this image is $\sim$ 25\arcmin. This image is
from \cite{Sij93}.
\label{minih}
}
\end{figure}

The diffuse extended cluster radio sources are currently grouped in
three classes: radio {\it halos}, {\it relics} and {\it mini-halos}.
A typical example of a cluster radio halo is shown in
Fig. \ref{a2163rx}.  The radio halos are permeating the cluster
central regions, with typical extent  \gtsim~1 Mpc and steep
spectrum.  Limits of a few percent to their polarized emission have been
derived. The  relic sources are similar to the halos in their low surface
brightness, large size and steep spectrum, but they are typically
found in the cluster peripheral regions  (see Fig.  \ref{a3667rx}).
Unlike halos, relics are highly polarized ($\sim$ 20\%).  
The mini-halos are detected around a dominant powerful radio galaxy at the
center of cooling core clusters, for a total size of $\sim$ 500 kpc,
as in the Perseus cluster (Fig. \ref{minih}).

The importance of the diffuse cluster sources is that they are large
scale features which are related to other cluster properties in the
optical and X-ray domain and are thus directly connected to the
cluster history and evolution.  These radio sources are difficult to
detect because of their low radio surface brightness and steep
spectrum.  Future generation instruments, as the SKA, will be crucial
 to study these sources at multiple frequencies and in polarization,
and to establish how common these ICM non-thermal components are in all
clusters.

Another interesting aspect of the ICM emission is represented by the
{\it Thomson scattering halos} produced at the center of cooling core
clusters where a strong radio galaxy is present.  These halos have not
been detected so far, and thus represent an important target for the
SKA observations.

\section{DIFFUSE SYNCHROTRON EMISSION}

The properties of large-scale radio halos and relics are poorly known,
because of the present observational limits.  Due to synchrotron and
inverse Compton losses, the typical lifetime of the relativistic
electrons in the ICM is relatively short ($\sim$ 10$^8$ yr), making it
difficult for the electrons to diffuse over a Mpc-scale region
within their radiative lifetime. The expected diffusion velocity of
the electron population is indeed of the order of the Alfv\'en speed,
$\sim$ 100 km/s.  Because diffuse sources extend throughout the
cluster volume, their electrons cannot have been injected or
reaccelerated in some localized points of the cluster, such as an
active galaxy or a shock, but they need in situ reacceleration.

\begin{figure}
\includegraphics[height=14pc]{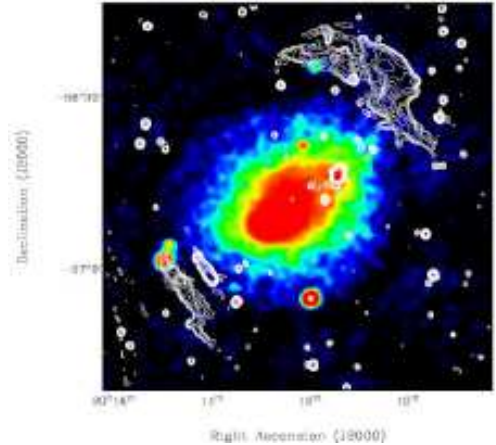}
\caption
{The cluster A3667 in radio and X-ray: the contours represent the
radio emission at 843 MHz (from \cite{Rot97}). The color scale
represents the ROSAT X-ray image emission. Two radio relics are located
on opposite sides of the cluster along the axis of the merger, with
the individual radio structures elongated perpendicular to this axis.
\label{a3667rx}
}
\end{figure}

Several suggestions for the mechanism transferring energy into the
relativistic electron population and for the origin of relativistic
electrons themselves have been made. Current models for radio halos
can be grouped in
two main classes, involving {\it primary electrons} and {\it secondary
electrons}, respectively \cite{B03}.  According to the first scenario, the
primary electrons were injected in the cluster volume from AGN
activity (quasars, radio galaxies, etc.), or from star formation in
normal galaxies (supernovae, galactic winds, etc.) during the cluster
dynamical history.  This population of electrons needs to be
reaccelerated \cite{P01} to compensate for the radiative losses.  It is
argued \cite{Fer03} that recent cluster mergers are likely to supply
energy to the halos, through stochastic turbulent acceleration 
\cite{Bru01,Fuj03} or shock acceleration, although the efficiency
of the last process is debated \cite{GB03,Ryu03}.  
The second class of models involves secondary electrons resulting from
inelastic nuclear collisions between the relativistic protons and the
thermal ions of the ambient intracluster medium.  The protons diffuse
on large scales because their energy losses are negligible.  They can
continuously produce in situ electrons, distributed through the
cluster volume \cite{BC99,Min01}.  In this scenario, it is
difficult to explain the observed association between mergers and
radio halos, the spectral index radial steepening found in Coma C and
the relatively low number of clusters with halos. However, the
secondary model can ultimately be tested with future gamma ray
telescopes due to its clear prediction of $\pi^0$-decay gamma ray
fluxes.

Different models have been suggested for the origin of the
relativistic electrons radiating in the radio relics, i.e.  located in
confined peripheral regions of the clusters.  There is increasing
evidence that the relics are tracers of shock waves in merger
events. This is consistent with their elongated structure almost
perpendicular to the merger axis (Fig. \ref{a3667rx}).  Active radio
galaxies may fill large volumes in the ICM with radio plasma, which
becomes rapidly invisible to radio telescopes because of radiation
losses of the relativistic electrons.  These patches of fossil radio
plasma are revived by adiabatic compression in a shock wave produced
in the ICM by the flows of cosmological large-scale structure
formation \cite{Enn98,EGK01}.

Unlike radio halos and relics, mini-halos are typically found at the
centers of cooling core clusters and are thus not connected to recent
cluster mergers.  Although these sources are generally surrounding a
powerful central radio galaxy, it has been argued \cite{GBS02} that
the energetic necessary to their maintenance is not supplied by the
radio galaxy itself. Current models involve electron reacceleration
by MHD turbulence in the cooling core region \cite{GBS02} or a hadronic
origin of the relativistic electrons from the interaction of cosmic
ray protons with the ambient thermal protons \cite{PE03}.

The investigation of diffuse cluster sources and the discrimination
between theoretical models is of great importance to the knowledge of
the formation and evolution of clusters of galaxies.

\section{RELATIVISTIC PLASMA AND CLUSTER MERGERS}

The study of radio halos and relics is directly connected to the
recent dynamical activity of clusters. In any scenario, cluster radio
halos give us deep insight into the physics and properties of galaxy
clusters.  This opens a new window of investigation of the properties
of clusters, through the formation of their relativistic components
and the connection between thermal and relativistic plasma.  Very
likely radio halos give a unique probe of non-thermal processes
accompanying energetic cluster merger events.

The typically low surface brightness (\ltsim $\mu$Jy arcsec$^{-2}$ at
1.4 GHz) of cluster radio halos and their steep spectrum makes it
difficult to image them accurately with the current resources.
Further, at lower resolution, where beam averaging enhances the
detectability of extended radio emission, true diffuse emission is
sometimes difficult to distinguish from a blend of weak, discrete
radio sources. Ideally, one wants high sensitivity on all angular
scales.

Current maps of cluster halos often miss some of the extended
structure, or show negative bowls around the imaged structure, arising
from missing short spacings.  Proper maps of cluster radio halos and
relics are only obtained with high sensitivity to low surface brightness,
i.e. a very good sampling of short baselines.

Multifrequency imaging is needed to derive spectra of these sources
and spectral index maps. Low frequency ($<$ 300 MHz) spectra are
important to determine the index of electron energy distribution,
while the high frequency spectra (up to 10 GHz) give information on
the diffusion and aging of relativistic particles, and any reacceleration
process.

Spectral index maps represent a powerful tool to study the properties
of the relativistic electrons and of the magnetic field in which they
emit, and to investigate the connection between the electron energy
and the ICM.  By combining high resolution spectral information and
X--ray images it is possible to study the thermal--relativistic plasma
connection both on small scales (e.g. spectral index variations
vs. clumps in the ICM distribution) and on large scales
(e.g. radial spectral index trends).  It has been shown \cite{Bru01}
that a relatively general expectation of models invoking
reacceleration of relic particles is a radial spectral steepening in
the synchrotron emission from radio halos.  Because of the low
diffusion velocity of the relativistic particles, the radial spectral
steepening cannot be simply due to aging of radio emitting
electrons. Therefore the spectral steepening must be related to the
intrinsic evolution of the local electron spectrum and to the radial
profile of the cluster magnetic field.  In this framework, radio
spectral index maps can be used to derive the physical conditions
prevailing in the clusters, i.e. reacceleration efficiency and 
magnetic field strength. Nowadays, only for three
clusters (Coma, A665, A2163) a spectral index image has been presented
in the literature, with resolutions of the order of 1\arcmin, and only
between two frequencies \cite{Giov93,Fer04}. 
Big developments are thus expected in this
area with SKA, which will allow multifrequency spectral studies
on higher resolution.

Polarization information on radio halos is not currently available,
because of the limited sensitivity. Polarimetric studies are important
to derive: i) direct information on the structure of
magnetic fields related with the cluster intergalactic medium,
ii) the magnetic field degree of ordering, related with
the evolution of radio emitting plasma.

SKA will dramatically improve the knowledge of radio halos, thanks to
spectral index studies and the detection of their polarized emission.

\section{SEARCH FOR NEW HALOS AND RELICS}

\begin{figure}
\includegraphics[scale=0.5]{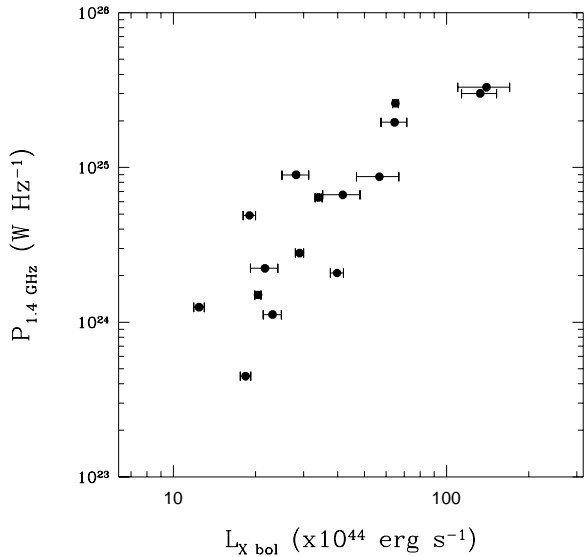}
\caption
{Halo monochromatic radio power at 1.4 GHz versus the cluster
bolometric X-ray luminosity for halos with size $>$1 Mpc.
\label{prlx}
}
\end{figure}

It is not yet clear if the ICM in every cluster
has a strong relativistic component.  The number of cluster sources of
this class is presently around 50, the most distant being in CL
0016+16 \cite{GF00} at z = 0.5545.  Most of them have been found owing
to searches in the NRAO VLA Sky Survey \cite{GTF99}, in the Westerbork
Northern Sky Survey \cite{KS01} and in the survey of the Shapley
Concentration \cite{Ven00}.  In a complete X-ray flux limited cluster
sample, the percentage of clusters showing diffuse radio sources at
the sensitivity limit of the NRAO VLA Sky Survey is 11\% (5\% halos
and 6\% relics) \cite{GF02}.  The detection rate increases up to
$\sim$35\% if only the clusters with X-ray luminosity larger than
10$^{45}$ erg s$^{-1}$ (in the ROSAT band 0.1-2.4 keV) are considered.
A correlation has been found between the radio power of a halo and the
X-ray luminosity of the parent cluster (Fig. \ref{prlx}). The
information on low power radio halos is currently unavailable.
Improved sensitivity to surface brightness will be crucial to find out
new cluster halos, to test if the above correlation can be extrapolated
to low radio powers and low X-ray luminosities, to study their
properties on a larger sample, and to test models.  Questions to answer are:
Do all clusters have a radio halo at some level? Do all clusters with
a recent merger have halos and relics?

Large numbers of galaxy clusters are expected to be found up to high
redhisfts by future surveys: e.g. the XMM Large Scale Structure
Survey is expected to find $\sim$ 10$^3$ galaxy clusters up to redshift
$\sim$ 1, SZ effect cluster detection with the Planck satellite
should find $\sim$ 10$^4$ galaxy clusters and the Sloan Digital 
Sky Survey is expected to identify $\sim$ 5 10$^5$ clusters.

Using radio halos and relics as tracers of cluster mergers 
will therefore allow investigations of the properties of the merger shocks
and turbulence in the accompanying clusters, and
detailed studies of the  cluster formation
process up to high resdhift.

In Table 1 we give the estimates of the number of observable halos
obtained by \cite{ER02}. These predictions are based on i) the
fraction of clusters containing radio halos, ii) the local X-ray
luminosity function and its evolution towards higher redshift, iii)
the local relation between X-ray cluster luminosity and radio halo
luminosity.  Estimates of the number of
relics were obtained by \cite{Brugg03} under the assumption that they
are produced via compression of old radio plasma. It was found that,
assuming a spectral index of 1, the number of radio relics and halos
roughly match at the same frequency. The possible detection of roughly
symmetric halos, located on opposite sides of a cluster as in A3667
(see Fig. \ref{a3667rx}), is particularly interesting as it could help
in understanding the geometry of the cluster formation process.

\begin{table}
   \caption{Estimates of the number of expected radio halos on the
full sky, which are above a given flux density S$_{tot}$ at 1.5 GHz, in
all clusters and in distant clusters(from \cite{ER02}). For the
details of the assumptions made for these calculation see the original
paper\cite{ER02}.  }
\medskip\medskip
   \begin{center}

 \begin{tabular}{l|c|c}
        \hline

  S$_{tot}$  & N  & N(z$>$0.3)  \\
         \hline \hline
  1 $\mu$Jy & 23759  & 10785 \\
  10 $\mu$Jy & 6812  & 2123 \\
  0.1 mJy &  1654 & 281 \\
  1 mJy & 326  & 21 \\
  10 mJy & 50  &  1\\
         \hline\hline

 \end{tabular}
 \end{center}
 \label{tab:num}
\end{table}

\section{RELEVANCE OF STUDIES WITH SKA}

The importance of the study of extended radio sources associated with
the ICM can be summarized in the following points.  The radio halos
are indicators of cluster mergers, tracers of high energy and/or
non-equilibrium plasma processes, probes of the ICM magnetic fields;
they will eventually allow to constrain models of
decaying/annihilating dark matter species. The radio relics are
tracers of shock waves occurring during the structure formation, they
will allow to determine shock properties like Mach numbers through
spectral index studies, and help to
clarify the nature of the seed relativistic
particle populations.  The radio mini-halos will allow to investigate
the interaction between the relativistic plasma and thermal plasma in
the cluster centers, and shed new light on the complex phenomena
occurring in the cluster cooling cores.  Moreover, detection of diffuse
sources at different redshifts will provide a detailed study of the
cluster formation process.

\begin{figure}
\includegraphics[height=18pc]{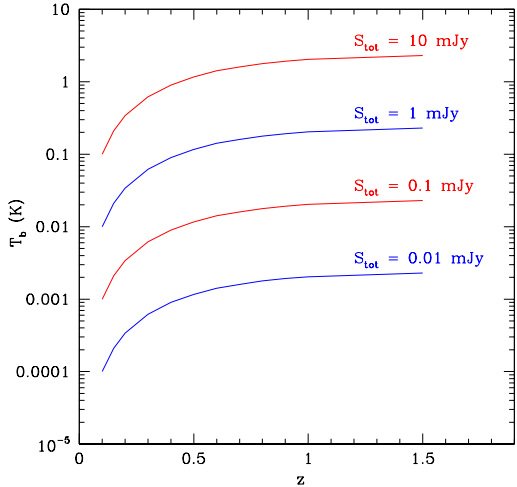}
\caption{ Brightness temperature at 1.4 GHz expected for radio halos
of a given total flux, as a function of redshift.  The size of a radio
halos is assumed of 1 Mpc.
\label{halodet}
}
\end{figure}

For the detection of extended diffuse radio sources in clusters, a
radio telescope must have high sensitivity to surface brightness, thus
an array with a very good coverage of short spacings.  In
Fig. \ref{halodet}, we show the estimates of the 1.4 GHz brightness of
radio halos of different total fluxes, computed by assuming a total
halo size of 1 Mpc.  With 50\% of the SKA collecting area within 5 km,
the brightness sensitivity at 1.4 GHz, with a beam of $\sim$ 8\arcsec,
is T$_b$ $\sim$ 5 mK (3$\sigma$ level), assuming a bandwidth of 0.5
GHz and an integration time of 1h.  This will allow to detect halos of
total flux down to 1 mJy at any redshift, and down to 0.1 mJy
at high redshift. Indicatively, a survey of 2$\pi$ sterad, would
detect about 300 new halos, about half of them at redshift $>$ 0.3 (see
Table 1). A similar number of relics will also be detected.  Of
course, the sensitivity will be much higher in deeper exposures on
targeted clusters, allowing detailed studies of the total and
polarized emission, at multiple frequencies.

A minimum baseline of 20 m allows detection of structures up to
34\arcmin~ at 1.4 GHz and up to 10\arcmin~ at 5 GHz.

Whereas the detection of diffuse emission needs an intermediate 
angular resolution, 
the high resolution of the full SKA is needed to resolve and
subtract discrete unrelated sources.

\section{THOMSON SCATTERING OF A CLUSTER CENTRAL RADIO SOURCE}

Although the hydrodynamical state and the ultimate fate of the cooling
gas is still uncertain, the central regions of the clusters with
cooling cores host large column densities of ionized material, as
directly derived by the observations.  These regions are likely to be
optically thin, with typical values of the optical depth due to
electron scattering of $\sim$10$^{-2}$.  If a powerful radio galaxy
resides at the center of a cooling core cluster, then diffuse
radiation originally emitted by the active nucleus and scattered by
the hot electrons in the cooling core region is expected to be
present.

The Thomson scattering in cluster cores has been described by \cite{Sun82}
and \cite{WS90}.  Since electron scattering is independent of photon
frequency, scattered radiation may in principle be detectable in any
part of the spectrum. However, it is unlikely that such faint diffuse
emission could be detected in the presence of other significant
emissions. Radio observations seem to be very suitable to observe this
effect, particularly since a very high dynamic range is attainable.

A diagnostic of the scattered radiation is the polarization.  This
emission should be characterized by a high degree of linear
polarization, with tangential orientation, and the polarization would
increase with the distance from the cluster center.

\begin{figure}
\includegraphics[width=8cm]{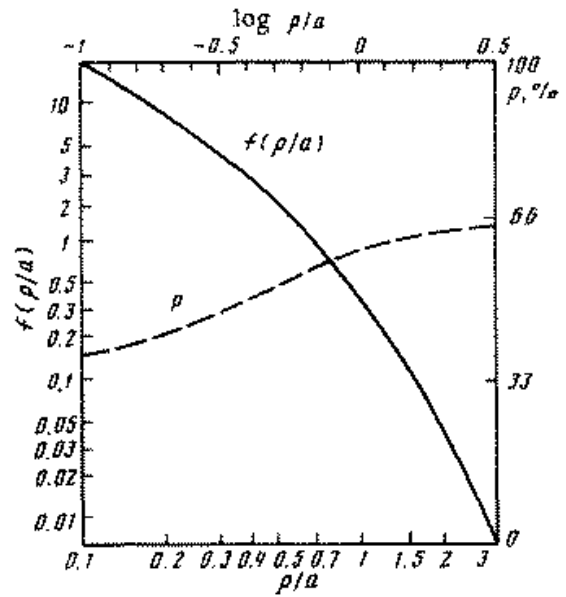}
\caption
{Solid curve: function describing the dependence of the diffuse
scattered radiation intensity on the projected distance 
$\rho$ from the compact radio source at the center of a
cluster of core radius $a$. Dashed curve: percentage polarization of
the diffuse radiation as a function of $\rho$/$a$ (from \cite{Sun82}).
\label{sunyaev}
}
\end{figure}

In the simple case that the cluster gas follows a King model, the
intensity of the scattered radiation and the amount of polarization
have the trends shown in Fig. \ref{sunyaev}.
The distribution of the brightness
temperature is given by the equation (from \cite{Sun82}):


\begin{eqnarray}
T_b = 0.92 \left({{S_\nu} \over {10~{\rm Jy}}}\right) \left({{\lambda} \over
{6~{\rm cm}}}\right)^2 \left({{10^{\prime}} \over {\theta_c}}\right) \nonumber \\
\left({{\tau_T} \over {10^{-2}}}\right)
{\rm f}({{\theta} \over {\theta_c}}) \ \ \ \ \ \ {\rm mK} \nonumber
\label{sunya}
\end{eqnarray}

\noindent
where $S_\nu$ denotes the monochromatic flux density, $\theta$ is the
scattering angle (see the original paper \cite{Sun82} for an accurate
definition of the geometry of the scattering), $\theta_c$ is the
angular core radius of the cluster, $\tau_T$ is the maximum optical
depth of the cluster, the function f(${{\theta} \over {\theta_c}}$) = f
(${{\rho} \over {a}}$) is plotted in Fig. \ref{sunyaev}.  It is estimated
that about 1\% of the luminosity of the cluster central source will be
scattered (see also \cite{WS90}).

Although the cluster gas approximation with a King model is too
simplified, we can use the above formula to get an order of magnitude
estimate of the intensity of the Thomson scattering effect.  We obtain
that the brightness temperature of the scattered radiation from a
source of 1 Jy at 1.4 GHz is of the order of $\approx$ 10 mK at the
center, and falls off very rapidly with distance (see
Fig. \ref{sunyaev}).

In a full-polarization study of the Perseus cluster in the mid-1990's,
de Bruyn (unpublished) found weak polarized emission.  They suggested
the possibility that this emission could be due to Thomson scattering.
Preliminary results from 
92 cm are presented by \cite{BdB03}. Beside the expected Galactic
Faraday depths around 8 rad m$^{-2}$, they found polarized emission
through the entire clusters at higher Faraday depth ($\sim$ 25 - 90
rad m$^{-2}$).  The simulations obtained to estimate the magnitude of
the effect in the Perseus cluster indicate that the observed emission
is too bright and too uniform in intensity to be explained by Thomson
scattering.

The detectability of the Thomson scattering halos will be possible with
new generation very sensitive instruments, with high sensitivity to
the polarized flux. An observational program designed to detect this
effect should include the study of radio point sources of
high-moderate strength associated with cluster-dominant galaxies in
cooling core clusters.  Since the scattered light is merely the 
emission of the central source redirected toward the observer, the
diffuse scattered component will have the same spectra index as the
central source.  Thus, given data of sufficient dynamic range and
spectral coverage, it should be possible to distinguish between
scattered radio photons and those from other sources.

By comparing the X-ray emission and electron scattering in a cooling
core, a distance to the cluster can be determined independently of the
value of the Hubble constant \cite{WS90}.  In addition, observations
of scattered light can provide information both on the central source
and the surrounding ICM. Time variability of the
central nucleus, beaming, or polarized emission all produce
characteristic profiles for the scattered emission which can be used
to study these processes.  Deviations from spherical symmetry or
clumping in the cooling flow itself can also produce characteristic
variations in the scattered surface brightness profile, allowing to
understand in an independent way some details of the cooling flow
process.

\vskip 1cm

{\bf ACKNOWLEDGMENTS} We are grateful to 
G. Brunetti, G. Giovannini and G. Setti for 
many interesting discussions.

\end{document}